\documentclass{llncs}
\usepackage[top=3cm,bottom=3cm,left=2.8cm,right=2.8cm]{geometry}
\usepackage[utf8]{inputenc}
\usepackage{fontawesome}
\usepackage{amsmath, amssymb}
\usepackage{stmaryrd}
\usepackage{mathtools}
\usepackage{csquotes}
\usepackage{xfrac}
\usepackage{xspace}
\usepackage{etoolbox}
\usepackage{xspace}
\usepackage[inline]{enumitem}
\usepackage[english]{babel}
\usepackage[usenames,dvipsnames]{xcolor}
\usepackage[hyperfootnotes=true,colorlinks,linkcolor=RedViolet,anchorcolor=black,citecolor=blue,urlcolor=blue]{hyperref}
\usepackage{todonotes}
\usepackage{caption}
\usepackage{palatino}

\newcommand{\codify}[1]{\textup{\texttt{\textbf{#1}}}}

\newcommand{\progvar}[1]{#1}
\newcommand{\assign}[2]{\progvar{#1} \coloneqq  {#2}}
\newcommand{\pchoice}[3]{\{{#1}\}[{#2}]\{{#3}\}}
\newcommand{\IF}[1]{\codify{if}({#1})\xspace}
\newcommand{\ELSE}{\codify{else}\xspace}
\newcommand{\WHILE}[1]{\codify{while}({#1})\xspace}
\newcommand{\observe}[1]{\codify{observe}({#1})}

\newcommand{\poisson}[1]{\mathtt{poisson}({#1})}

\newcommand{\toolname}[1]{\textsc{#1}\xspace}

\makeatletter
\renewcommand{\fnum@figure}{Prog.~\thefigure}
\makeatother

\newif\ifanonymous
\anonymousfalse

\newcommand{\anonymize}{\ifanonymous Anon.\ Author(s) \else Lutz Klinkenberg \and Tobias Winkler \and Mingshuai Chen \and Joost-Pieter Katoen\fi}
\newcommand{\anonymizeinst}{\ifanonymous \else RWTH Aachen University, Aachen, Germany \\
\email{\{lutz.klinkenberg,tobias.winkler,chenms,katoen\}@cs.rwth-aachen.de} \fi}
\newcommand{\anonymizelink}[1]{\ifanonymous Link omitted due to blind reviews \else \faGithub~\url{#1} \fi.}

\title{Exact Probabilistic Inference Using Generating Functions\thanks{Extended abstract accepted by LAFI 2023 -- the Languages for Inference workshop co-located with POPL 2023.}}
\author{\anonymize}
\institute{\anonymizeinst}

\patchcmd{\paragraph}{\itshape}{\bfseries\boldmath}{}{} 

\begin{document}

\maketitle

\setlength{\floatsep}{1\baselineskip}
\setlength{\textfloatsep}{1\baselineskip}
\setlength{\intextsep}{1\baselineskip}

Probabilistic programs are typically normal-looking 
programs describing posterior probability distributions.
They intrinsically code up randomized algorithms and have long been at the heart of modern machine learning and approximate computing. 
We explore the theory of \emph{generating functions}~\cite{generatingfunctionology} and investigate its usage in the exact quantitative reasoning of 
probabilistic programs.
Important topics include the exact representation of program semantics~\cite{LOPSTR}, proving exact program equivalence~\cite{CAV22}, and -- as our main focus in this extended abstract -- exact probabilistic inference. 

In probabilistic programming, \emph{inference} aims to derive a program's posterior distribution.
In contrast to approximate inference, inferring \emph{exact} distributions comes with several benefits~\cite{DBLP:conf/pldi/GehrSV20}, e.g., no loss of precision, natural support for symbolic parameters, and efficiency on models with certain structures.
Exact probabilistic inference, however, is a notoriously hard task~\cite{DBLP:journals/ai/Cooper90,DBLP:journals/acta/KaminskiKM19,DBLP:journals/toplas/OlmedoGJKKM18,ROTH1996273}.
The challenges mainly arise from three program constructs:
\begin{enumerate*}[label=(\arabic*)]
    \item\label{issue:loop} unbounded while-loops and/or recursion,
    \item\label{issue:support} infinite-support distributions, and 
    \item\label{issue:conditioning} conditioning (via posterior observations).
\end{enumerate*}
We present our ongoing research in addressing these challenges (with a focus on conditioning) leveraging generating functions and show their potential in facilitating exact probabilistic inference for discrete probabilistic programs.
%

\section{Inference in Probabilistic Programs}

\paragraph{State-of-the-Art.}
Most existing probabilistic programming languages implement \emph{sampling}-based inference algorithms rooted in the principles
of Monte Carlo \cite{doi:10.1080/01621459.1949.10483310}, thereby yielding numerical approximations of the exact results, see, e.g., \cite{gram2021extending}.
In terms of semantics, many probabilistic systems employ \emph{probability density function} (PDF) representations of distributions, e.g., ($\lambda$)\toolname{PSI}~\cite{DBLP:conf/cav/GehrMV16,DBLP:conf/pldi/GehrSV20}, \toolname{AQUA}~\cite{DBLP:conf/atva/HuangDM21}, \toolname{Hakaru}~\cite{DBLP:conf/flops/NarayananCRSZ16}, and the density compiler in~\cite{DBLP:conf/popl/BhatAVG12,DBLP:journals/lmcs/BhatBGR17}. These systems are dedicated to 
inference (with conditioning) for programs encoding (joint discrete-)continuous distributions.
Reasoning about the underlying PDF representations, however, amounts to resolving complex integral expressions in order to answer inference queries, thus confining these techniques either to (semi-)numerical methods \cite{DBLP:conf/popl/BhatAVG12,DBLP:journals/lmcs/BhatBGR17,DBLP:conf/atva/HuangDM21,DBLP:conf/flops/NarayananCRSZ16} or exact methods yet limited to bounded looping behaviors \cite{DBLP:conf/cav/GehrMV16,DBLP:conf/pldi/GehrSV20}. \toolname{Dice}~\cite{DBLP:journals/pacmpl/HoltzenBM20} employs weighted model counting to enable exact inference for discrete probabilistic programs, yet is also confined to statically bounded loops. The tool \toolname{Mora}~\cite{DBLP:conf/ictac/BartocciKS20,DBLP:conf/tacas/BartocciKS20} supports exact inference for various types of Bayesian networks, but relies on a restricted form of intermediate representation known as prob-solvable loops.

\paragraph{The PGF Approach.}
Klinkenberg et al.~\cite{LOPSTR} provide a program semantics that allows for exact quantitative reasoning about probabilistic programs without conditioning.
They exploit a denotational approach \`{a} la Kozen~\cite{Kozen} and treat a probabilistic program as a \emph{distribution transformer}, i.e., mapping a distribution over the inputs (the prior) into a distribution after execution of the program (the posterior).
In \cite{LOPSTR}, the domain of discrete distributions is represented in terms of \emph{probability generating functions} (PGFs), which are a special kind of generating functions~\cite{generatingfunctionology}.
This representation comes with several benefits:
\begin{enumerate*}[label=(\alph*)]
    \item it naturally encodes common, infinite-support distributions (and variations thereof) like the geometric or Poisson distribution in compact, \emph{closed-form} representations;
    \item it allows for compositional reasoning and, in particular, 
    in contrast to representations in terms of density or mass functions, the effective computation of (high-order) moments;
    \item tail bounds, concentration bounds, and other properties of interest can be extracted with relative ease from a PGF; and
    \item expressions containing parameters, both for probabilities and for assigning new values to program variables, are naturally supported.
\end{enumerate*}
Some successfully implemented ideas based on PGFs, e.g., for deciding probabilistic equivalence and for proving non-almost-sure termination, are presented in~\cite{CAV22,LOPSTR}, which address especially the aforementioned challenges \ref{issue:loop} and \ref{issue:support} for exact probabilistic inference \emph{without} conditioning.

\section{Taming Conditioning Using PGFs}
The creation of generative models is a challenging task, as these models oftentimes need expert domain knowledge.
Therefore, the concept of \emph{conditioning as a first-class language element} is crucial as it allows for a natural and intuitive approach to the creation of models.
Our current research aims to \emph{extend the PGF approach towards exact inference for probabilistic programs with conditioning -- thus addressing challenges \textnormal{\ref{issue:loop}}, \textnormal{\ref{issue:support}}, and \textnormal{\ref{issue:conditioning}} -- and to push the limits of automation as far as possible}. To this~end, we are in the process of developing an exact, symbolic inference engine based on the open-source, PGF-based tool \toolname{Prodigy} \cite{CAV22}.
We illustrate below its current capability to cater for conditioning via two examples.
%
\vspace*{-\baselineskip}
\begin{figure}[h]
\begin{minipage}[b]{0.5\textwidth}
    \begin{align*}
        & \pchoice{\assign{w}{0}}{\sfrac{5}{7}}{\assign{w}{1}}\fatsemi \\[.1cm]
        & \IF{\progvar{w} = 0} \{\ \assign{c}{\poisson{6}}\ \}\\
        & \ELSE\ \{\ \assign{c}{\poisson{2}}\ \}\fatsemi \\[.2cm]
        & \observe{\progvar{c} = 5}
    \end{align*}
    \captionof{figure}{The telephone operator.}
    \label{fig:toperator}
\end{minipage}
\begin{minipage}[b]{0.5\textwidth}
    \begin{align*}
        & \assign{x}{1}\fatsemi\\[.1cm]
        & \WHILE{\progvar{x} = 1} \{\\
        & \quad \pchoice{\assign{c}{\progvar{c}+1}}{\sfrac{1}{2}}{\assign{x}{0}}\fatsemi\ \}\\[.2cm]
        & \observe{\progvar{c} \equiv 1 \!\!\!\!\pmod{2}}
    \end{align*}
    \captionof{figure}{The odd geometric distribution.}
    \label{fig:loopyprog}
\end{minipage}
\end{figure}
\vspace*{-\baselineskip}
\paragraph{\it Conditioning in Loop-Free Programs.}
Prog.\ \ref{fig:toperator} is a loop-free probabilistic program encoding an infinite-support distribution.
It describes a telephone operator who is unaware of whether today is a weekday or weekend.
The operator's initial belief is that with probability $\sfrac{5}{7}$ it is a weekday ($\progvar{w}=0$) and thus with probability $\sfrac{2}{7}$ weekend ($w=1$).
Usually, on weekdays there are 6 incoming calls per hour on average;
on weekends this rate decreases to 2 calls -- both rates are subject to a Poisson distribution.
The operator observes 5 calls in the last hour.
The inference task is 
to compute the distribution in which the initial belief is updated based on the posterior observation.
\toolname{Prodigy} can automatically infer that $\textup{Pr}(\progvar{w}=0) = \frac{1215}{1215 + 2 \cdot e^4} \approx 0.9178$.

\paragraph{\it Conditioning Outside of Loops.}
Prog.\ \ref{fig:loopyprog} describes an iterative algorithm that repeatedly flips a fair coin -- while counting the number of trials -- until seeing heads, and observes that this number is odd. Whereas Prog.\ \ref{fig:toperator} can be handled by ($\lambda$)\toolname{PSI}~\cite{DBLP:conf/cav/GehrMV16,DBLP:conf/pldi/GehrSV20}, Prog.\ \ref{fig:loopyprog} cannot, as ($\lambda$)\toolname{PSI} do not support programs with unbounded looping behaviors.
However, given a suitable invariant as described in \cite{CAV22}, \toolname{Prodigy} is able to reason about the posterior distribution of Prog.\ \ref{fig:loopyprog} in an automated fashion using the \emph{second-order PGF} (SOP) technique \cite{CAV22}:
the resulting posterior distribution for any input with $\progvar{c}=0$ is $\frac{3 \cdot c}{(4-c^2)}$ which encodes precisely a closed-form solution for the generating function $\sum_{n=0}^{\infty} -3 \cdot c^n \cdot \left(2^{-2-n} \cdot \left(-1 + (-1)^n\right)\right)$.

\section{Future Directions}

A natural question is whether we can tackle exact inference when conditioning occurs \emph{inside} of a loop.
As argued in \cite{DBLP:journals/toplas/OlmedoGJKKM18}, more advanced inference techniques are required to answer this question. In fact, to the best of our knowledge, there is no (semi-)automated exact inference technique that allows for the presence of observe statements inside a (possibly unbounded) loop (an exception could be the potentially automatable conditional weakest preexpectation calculus \cite{DBLP:journals/toplas/OlmedoGJKKM18}).
This is precisely our current research focus.
One promising idea is to develop a non-trivial syntactic restriction of the programming language, where the more advanced SOP technique \cite{CAV22} can be generalized to address conditioning inside loops.

The possibility to incorporate symbolic parameters in PGF representations can enable the application of well-established optimization methods, e.g., maximum-likelihood estimations and parameter fitting, to the inference for probabilistic programs.
Other interesting future directions include deciding equivalence of probabilistic programs with conditioning, amending our method to continuous distributions using characteristic functions, and exploring the potential of PGFs in differentiable programming.
%


\bibliographystyle{plain}
\bibliography{references.bib}

\end{document}